\documentclass[journal]{IEEEtran}

\usepackage{cite}
\usepackage{amsmath,amssymb,amsfonts}
\usepackage{graphicx}
\usepackage{textcomp}
\usepackage{xcolor}
\usepackage{booktabs}
\usepackage{array}
\usepackage{bm}
\usepackage{balance}
\usepackage[colorlinks=true,linkcolor=blue,citecolor=blue,urlcolor=blue]{hyperref}

\newcommand{\vect}[1]{\bm{#1}}
\newcommand{\CRB}{\mathrm{CRB}}

\begin{document}

\title{MI-ISAC: Magneto-Inductive Integrated Sensing and Communication\\
in the Reactive Near-Field for RF-Denied Environments}

\author{Haofan~Dong,~\IEEEmembership{Student Member,~IEEE,}
  and~Ozgur~B.~Akan,~\IEEEmembership{Fellow,~IEEE}%
\thanks{H.~Dong and O.~B.~Akan are with the Internet of Everything Group,
Department of Engineering, University of Cambridge, Cambridge CB3~0FA, U.K.
(e-mail: hd489@cam.ac.uk; oba21@cam.ac.uk).}
\thanks{Ozgur B. Akan is also with the Center for neXt-generation Communications
(CXC), Department of Electrical and Electronics Engineering, Koç University, 34450 Istanbul, Turkey (email:oba21@cam.ac.uk)}}

\markboth{IEEE Wireless Communications Magazine, 2026}%
{Dong and Akan: Magneto-Inductive ISAC in the Reactive Near-Field}
\maketitle

\begin{abstract}
Radio-frequency integrated sensing and communication (RF-ISAC) is ineffective in
underground, underwater, and in-body environments where conductive media
attenuate electromagnetic waves by tens of dB per meter.
This article presents magneto-inductive ISAC (MI-ISAC), a paradigm that exploits the reactive near-field quasi-static coupling inherent to MI links, enabling a fundamentally different approach to ISAC in these RF-denied environments.
Five foundational results are established:
(i)~tri-axial coils are necessary and sufficient for identifiable
joint range-and-angle estimation;
(ii)~coupling strength changes sharply with range, enabling
theoretical sub-millimeter accuracy at typical MI distances despite kHz-level bandwidth;
(iii)~time-of-flight is ineffective under such narrow bandwidth, but
the coupling gradient provides approximately six orders of magnitude
finer resolution;
(iv)~MI-ISAC can provide 4--10+\,dB sensing gain over time-division
baselines; and
(v)~the MI-MIMO channel is geometry-invariant and well-conditioned
across all orientations.
Applications and a research roadmap are discussed.
\end{abstract}


\section{Introduction}
\label{sec:intro}

\IEEEPARstart{T}{he} vision of 6G wireless systems places integrated
sensing and communication (ISAC) at center stage: a single platform
simultaneously transmits data and perceives its physical
environment~\cite{liu2022survey,liu2022fundamental}.
Yet the existing ISAC literature is built upon radio-frequency (RF)
electromagnetic wave propagation---a premise that does not hold in three
domains.

\emph{Underground.}
Soil and rock absorb RF energy at 10--30\,dB/m (sub-GHz), restricting
conventional radar and communication to a few meters.
Mining rescue, pipeline integrity monitoring, precision agriculture, and
seismic early-warning networks all require coexisting communication and
localization capabilities that RF-ISAC cannot
provide~\cite{akyildiz2006wireless,saeed2019iot}.
In these scenarios, knowing \emph{where} an underground sensor sits is
as important as the data it reports.

\emph{Underwater.}
Seawater conductivity ($\approx$4\,S/m) limits RF to centimeters;
acoustic links impose $\sim$1\,s/km latency, hindering real-time
AUV control~\cite{akyildiz2015realizing}.

\emph{In-body.}
Biological tissue absorbs RF significantly, and SAR limits constrain power to milliwatts.
Next-generation implants need bidirectional telemetry coupled
with local tissue sensing for closed-loop therapy~\cite{ma2025survey}.

\begin{figure*}[!t]
  \centering
  \includegraphics[width=0.92\textwidth]{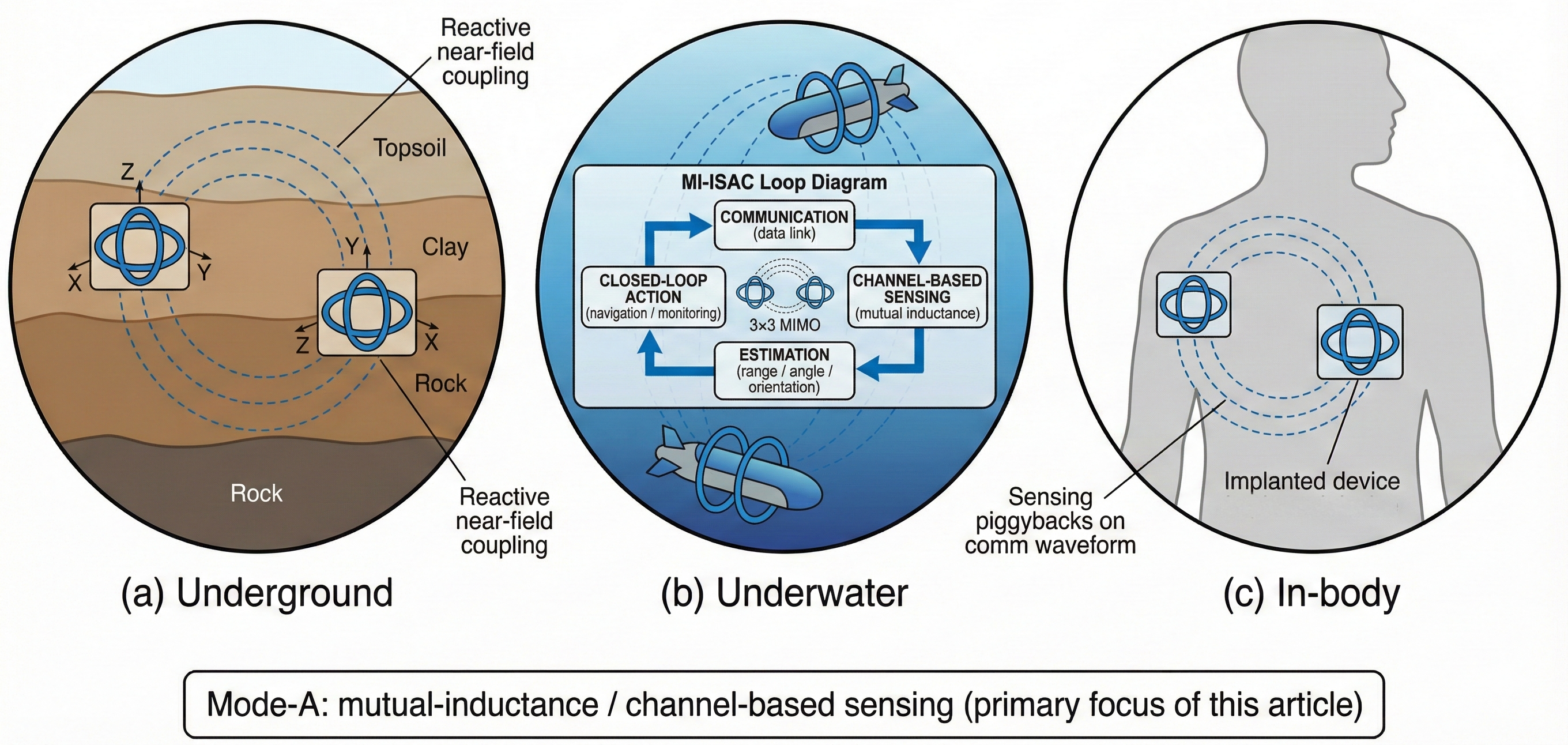}
  \caption{MI-ISAC concept: three RF-denied environments (underground,
    underwater, in-body) unified by a single reactive near-field
    coupling link.
    The central MI-ISAC loop illustrates how communication,
    channel-based sensing, parameter estimation, and closed-loop
    control share the same waveform and hardware through the
    geometry-deterministic $3\!\times\!3$ MI-MIMO channel.}
  \label{fig:concept}
\end{figure*}

All three domains require co-located communication and sensing---the ISAC
mandate---which RF-ISAC cannot adequately address (Fig.~\ref{fig:concept}).

Recent near-field ISAC work uses large arrays in the \emph{radiative}
Fresnel region~\cite{cui2023nearfield}, where sensing
still relies on ToF/Doppler with a spherical-wave refinement.
MI operates in the \emph{reactive} near field, where radiation is
negligible and interaction is dominated by quasi-static magnetic
coupling---so the physics, models, and design tools require substantial
revision.

This article introduces MI-ISAC as a new paradigm and presents five
quantifiable results:
\begin{enumerate}
\item \textbf{MI-ISAC framework:} MI-ISAC is defined formally and it is shown
  that, to the best of our knowledge, no prior work exists at the
  MI$\times$ISAC intersection, despite the IEEE COMST 2025 MI
  survey~\cite{ma2025survey} noting this as an open area.

\item \textbf{Resolution paradigm shift:}
  MI-ISAC ranging resolution is determined by the coupling gradient
  ($r^{-3}$ amplitude), not bandwidth; the RF ToF limit is inapplicable
  at MI bandwidths ($c/2B = 150$\,km at 1\,kHz).

\item \textbf{Identifiability and CRB:}
  Tri-axial coils yield a full-rank Fisher information matrix (FIM)
  for $(r,\theta,\phi)$; single-axis coils are rank-deficient.
  The CRB scales as $r^{8}$ in closed form.

\item \textbf{MI-MIMO structure:}
  The dipole coupling tensor $\vect{G}$ has eigenvalues
  $\{+2,-1,-1\}$, giving rank\,=\,3 and $\kappa = 2$ universally.

\item \textbf{Quantifiable ISAC gain:}
  MI-ISAC can achieve 4--10+\,dB sensing gain over a TDMA baseline,
  decomposed into a 3-dB time-multiplexing term and a 1--9\,dB
  structural-information term.
\end{enumerate}

\section{MI Communication Fundamentals}
\label{sec:primer}

\subsection{From Wave Propagation to Quasi-Static Coupling}
\label{subsec:reactive}

In conventional RF communication, information is carried by
electromagnetic waves that radiate from a transmitter and propagate
through space at the speed of light.
In MI communication, information transfer occurs through a different
mechanism: \emph{reactive near-field magnetic
coupling}~\cite{sun2010magnetic}.
A transmitter coil generates a quasi-static magnetic field; a receiver coil
experiences induced voltage proportional to mutual inductance.
Propagation delay is negligible---energy exchanges through flux linkage
rather than radiation.

The implication for sensing: with negligible propagation delay, the
received signal strength depends solely on geometry and the medium---precisely
the basis of MI-ISAC sensing.
Moreover, since $\mu \approx \mu_0$ in soil, seawater, and tissue,
MI path loss is largely medium-independent, unlike
RF~\cite{sun2010magnetic,gulbahar2012underwater}.

\subsection{The Deterministic MI Channel}
\label{subsec:channel}

Consider a Tx and Rx coil separated by distance $r$, each with
$N_t$ turns and radius $a$, operating at angular frequency
$\omega_{0} = 2\pi f_{0}$ (Fig.~\ref{fig:system_model}).
In the magnetic dipole approximation ($a \ll r \ll \lambda$), the
channel coefficient is:
\begin{equation}
  h(r,\theta,\phi) \;=\; \frac{C}{r^{3}}\;\cdot\; g(\theta,\phi),
  \label{eq:channel}
\end{equation}
with $C$ a known coil-dependent constant\footnote{$C = \mu_{0}\omega_{0} N_t^{2} A^{2} / (4\pi)$ with $A = \pi a^{2}$ the coil area; in practice $C$ is calibrated per deployment to absorb coil non-idealities.}
 and $g(\theta,\phi)$ encoding the orientation dependence through the dipole coupling tensor $\vect{G}$.
The received signal is the transmitted symbol scaled by the channel
coefficient plus additive white Gaussian noise.

Two properties of~\eqref{eq:channel} are central to MI-ISAC.
First, the channel is \emph{deterministic}: it exhibits negligible fading, multipath, and
stochastic variation---a known, invertible function of geometry.
Second, it has a \emph{pronounced gradient}: the $r^{-3}$ amplitude dependence creates a sensitivity that
scales as $r^{-4}$---a limitation for communication range but a significant
advantage for range estimation.

\subsection{Tri-Axial Coils and $3\!\times\!3$ MIMO}
\label{subsec:triaxial}

\begin{figure}[!t]
  \centering
  \includegraphics[width=\columnwidth]{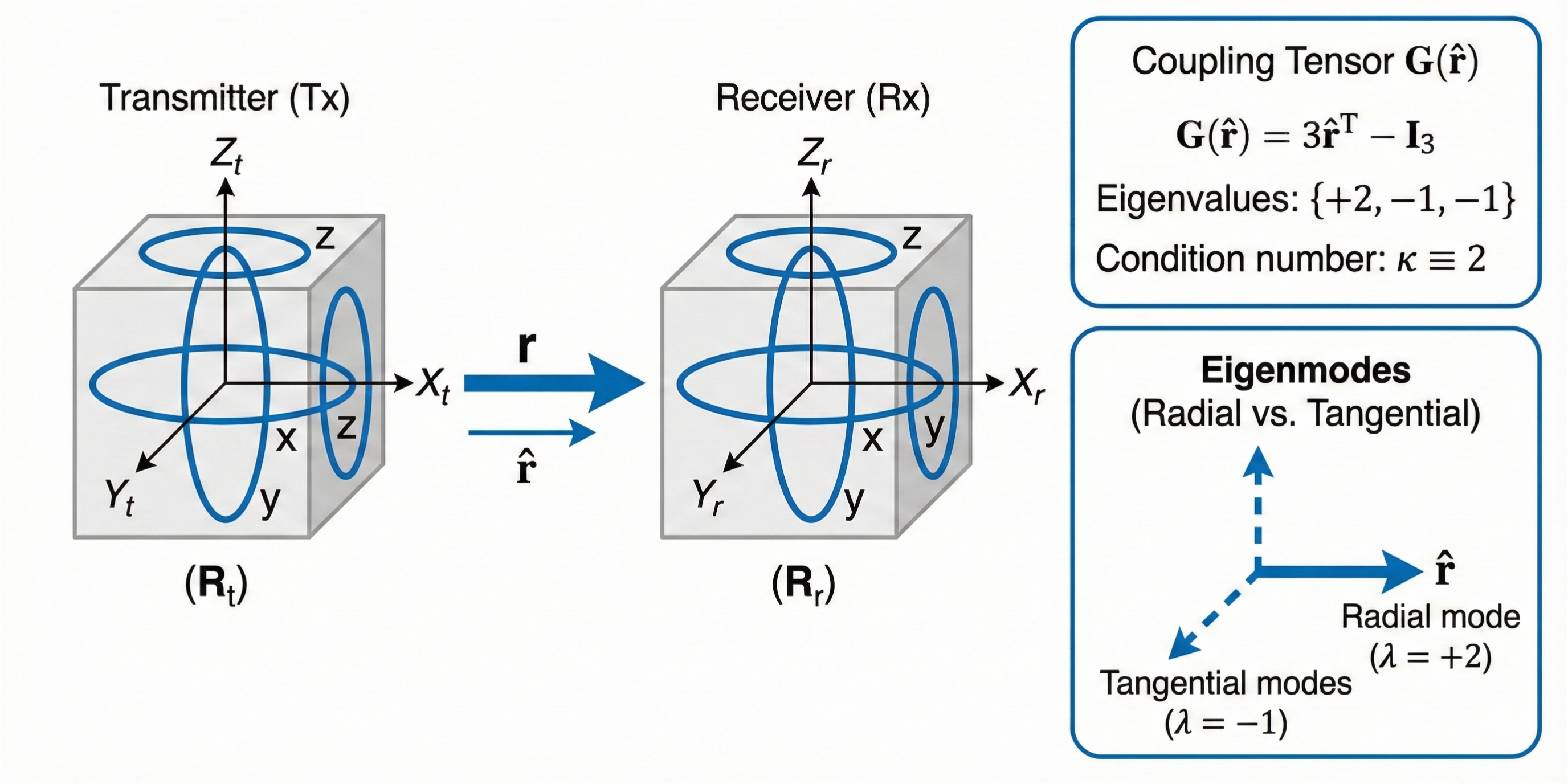}
  \caption{MI channel model and tri-axial coil geometry.
    Two nodes with local coordinate frames
    $(X_t,Y_t,Z_t)$ and $(X_r,Y_r,Z_r)$ are separated by vector
    $\vect{r}$.
    The coupling tensor
    $\vect{G} = 3\hat{\vect{r}}\hat{\vect{r}}^{\!\top}-\vect{I}_3$
    has eigenvalues $\{+2,-1,-1\}$ and corresponding
    radial/tangential eigenmodes.}
  \label{fig:system_model}
\end{figure}

A single-axis coil captures only the projection of the magnetic field
along its normal.
To recover the full vector field, each node is equipped with \emph{three
mutually orthogonal coils}---a tri-axial configuration---creating a $3\!\times\!3$ MI-MIMO channel whose entries are
governed by the dipole coupling tensor $\vect{G}$ (Fig.~\ref{fig:system_model}).
Tri-axial coil hardware is well-established: it has been demonstrated in
underground MI networks~\cite{kisseleff2014throughput}, magneto-inductive
positioning systems~\cite{markham2012miners}, and medical
telemetry~\cite{ma2025survey}.
As shown in Section~\ref{sec:results}, the tri-axial architecture is
not only beneficial but \emph{necessary} for identifiable
sensing: single-axis coils cannot uniquely determine all three spatial
parameters.

\section{MI-ISAC: Concept and System Architecture}
\label{sec:architecture}

\subsection{What Is MI-ISAC?}
\label{subsec:definition}

MI-ISAC is defined as follows:
\emph{a system in which the same set of MI communication symbols and
packet structure simultaneously supports both data transmission and
estimation of physical parameters $(r,\theta,\phi)$ or environmental
parameters $(\sigma_{m})$, using a single hardware platform without
dedicated sensing waveforms or additional spectral resources.}

This definition distinguishes MI-ISAC from prior work on MI-based
localization~\cite{saeed2019iot,lin2015cross,markham2012miners}, which
treated positioning and communication as separate functions using
distinct time slots or frequency bands.
The most relevant prior work, Lin \emph{et al.}~\cite{lin2015cross}, performed
sequential MI communication then MI localization---a partially
integrated design analogous to time-division ISAC.
The framework presented here targets \emph{full integration} at the waveform level.

Table~\ref{tab:comparison} contrasts RF-ISAC, MI-ISAC, and the prior MI
positioning-plus-communication approach across seven dimensions, making
the fundamental distinctions explicit.

\begin{table*}[!t]
  \centering
  \caption{Fundamental Comparison: RF-ISAC vs.\ MI-ISAC (Mode~A) vs.\
    Prior MI Positioning-plus-Communication}
  \label{tab:comparison}
  \footnotesize
  \begin{tabular}{@{} l l l l @{}}
    \toprule
    \textbf{Attribute}
      & \textbf{RF-ISAC}
      & \textbf{MI-ISAC (this work)}
      & \textbf{Prior MI pos.+comm.} \\
    \midrule
    EM regime
      & Radiative far/near-field
      & Reactive near-field
      & Reactive near-field \\
    Sensing mechanism
      & ToF / Doppler
      & Coupling strength \& tensor
      & RSS fingerprinting \\
    Channel model
      & Stochastic (fading)
      & Deterministic $h(r,\theta,\phi)$
      & Deterministic \\
    Integration level
      & Waveform-level
      & Waveform-level
      & Time-division (partial) \\
    MIMO structure
            & Array manifold; $\kappa$ depends on scattering
      & Dipole tensor $\vect{G}$, $\kappa\!\equiv\!2$
      & Single-axis typical \\
    Resolution source
      & Bandwidth $B$
      & Coupling gradient $r^{-3}$
      & Calibration database \\
    Target environment
      & LoS air / indoor
      & RF-denied media
      & RF-denied media \\
    \bottomrule
  \end{tabular}
\end{table*}

\subsection{Sensing Modes: Mode~A vs.\ Mode~B}
\label{subsec:modes}

MI-ISAC sensing can operate in two complementary modes.
\textbf{Mode~A (Bistatic, Link-Mediated)} extracts parameters from the
mutual-inductance channel $h(r,\theta,\phi,\sigma_m)$ between Tx and Rx.
Each communication symbol carries sensing information through its
amplitude and phase.
This is the primary mode addressed in this article---it enables low-overhead sensing
because the receiver already estimates the channel for
demodulation; inverting the deterministic channel model to obtain
geometric parameters requires only additional computation, no additional
signaling.
\textbf{Mode~B (Monostatic, Self-Impedance)} detects nearby conductors
via eddy-current-induced impedance changes~\cite{kraichman1962impedance},
but exhibits $r^{-6}$ decay limiting range to 5--10\,m; it is treated as a
future extension.

\subsection{MI-ISAC Transceiver Architecture}
\label{subsec:transceiver}

A minimal MI-ISAC transceiver generates standard communication frames:
a short preamble for synchronization followed by data symbols.
The Rx performs three parallel tasks:
synchronization and frequency tracking;
channel estimation followed by data demodulation (communication
path); and parameter extraction from the estimated channel (sensing
path).
In the sensing path, the main operation is \emph{channel inversion}:
the receiver extracts range from the estimated channel magnitude
and angles from the eigenstructure of the estimated coupling tensor.
Because $\vect{H}$ is a deterministic, bijective function of
$(r,\theta,\phi)$, this inversion is well-posed whenever the FIM is
full-rank---a condition guaranteed by the tri-axial architecture.
Since channel estimation is already performed for coherent demodulation,
the only added sensing cost is a lightweight algebraic
inversion---achieving sensing at negligible additional overhead.

\section{Key Results: Five Foundational Insights}
\label{sec:results}

\subsection{Insight~1: Tri-Axial Is Necessary for Identifiability}
\label{subsec:identifiability}

The fundamental question for any sensing system is: \emph{can the
parameters of interest be uniquely determined from observations?}
For MI-ISAC, the goal is to estimate range, azimuth, and elevation from received
signals. The Fisher information matrix (FIM) captures the available
information; identifiability requires the FIM to be full-rank.

Under narrowband MI mutual-inductance observations, a single-axis coil
yields a rank-deficient FIM, making joint
estimation of $(r,\theta,\phi)$ infeasible.
The intuition is straightforward: a single scalar observation $|h|$ couples $r$,
$\theta$, and $\phi$ through one equation---three unknowns from a single
equation, hence rank deficiency.
A tri-axial coil configuration at both Tx and Rx, however, yields a
full-rank FIM for all non-degenerate geometries, because the
$3\!\times\!3$ matrix observation $\vect{H}$ provides up to 9
independent real measurements (6 unique due to symmetry), more than
sufficient to resolve 3~parameters.
Fig.~\ref{fig:crb_fim}(a) illustrates this via the FIM rank comparison.
This implies that \emph{tri-axial coils are not only
beneficial but necessary for MI-ISAC.}
While the dipole tensor eigenstructure is established in
electromagnetics, its role as an \emph{identifiability enabler} for
joint communication-and-sensing is, to our knowledge, new.

\subsection{Insight~2: CRB Scales as $r^{8}$}
\label{subsec:crb}

Given identifiability, the next question is: \emph{how precisely can $r$
be estimated?}
Because the channel amplitude decays as $r^{-3}$, its gradient with
respect to range scales as $r^{-4}$.
Squaring this gradient in the Fisher information yields the
Cram\'{e}r--Rao bound:
\begin{equation}
  \CRB(r) \;=\; \frac{\sigma_w^{2}\; r^{\,8}}{18\, N\, P\, |C|^{2}},
  \label{eq:crb_r8}
\end{equation}
where $N$ is the number of observed symbols, $P$ the transmit power,
and $\sigma_w^{2}$ the noise variance.

\begin{figure}[!t]
  \centering
  \includegraphics[width=\columnwidth]{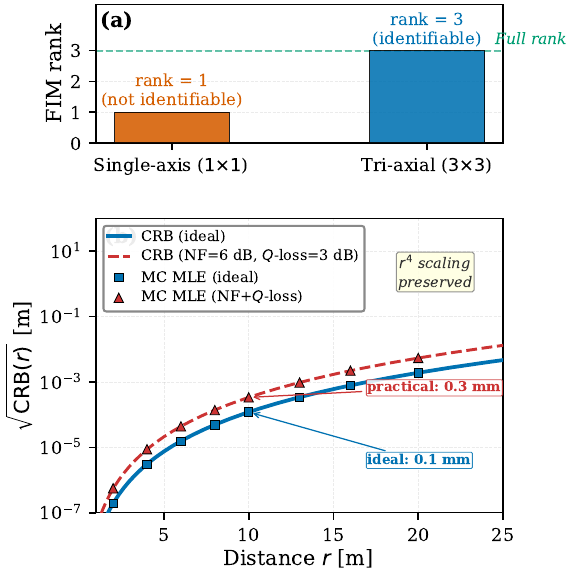}
\caption{\textbf{(a)}~FIM rank: single-axis (rank\,=\,1, not
  identifiable) vs.\ tri-axial (rank\,=\,3, identifiable).
  \textbf{(b)}~$\sqrt{\mathrm{CRB}(r)}$ vs.\ distance showing $r^{4}$
  scaling (equivalently $r^{8}$ in CRB).
  Blue: ideal thermal noise; red: practical front-end
  (NF\,=\,6\,dB, finite-$Q$ loss\,=\,3\,dB).
  MC~MLE markers validate both bounds.
  At $r\!=\!10$\,m: 0.1\,mm (ideal) vs.\ 0.3\,mm (practical)---both
  sub-millimeter.
  Parameters: $a\!=\!0.15$\,m, $N_t\!=\!20$, $f_0\!=\!10$\,kHz,
  $N\!=\!100$, $P\!=\!1$\,W, $B\!=\!1$\,kHz.}
  \label{fig:crb_fim}
\end{figure}

The $r^{8}$ scaling arises from the $r^{-3}$ coupling law.
While this implies rapid degradation at long range, at typical MI
communication distances (1--30\,m) it enables high precision in principle.
For representative underground parameters ($a\!=\!0.15$\,m,
$N_t\!=\!20$, $f_0\!=\!10$\,kHz, $N\!=\!100$, $P\!=\!1$\,W),
the root-CRB, representing the theoretical lower bound, at 10\,m is approximately 0.1\,mm under ideal thermal
noise---a theoretical limit of sub-millimeter ranging accuracy from a narrowband 10\,kHz link.
Fig.~\ref{fig:crb_fim}(b) validates the $r^{8}$ scaling via Monte
Carlo MLE that achieves the bound under both ideal and practical
front-end conditions (noise figure 6\,dB, finite-$Q$ insertion loss
3\,dB); the latter shifts the CRB upward by 9\,dB but preserves the
$r^{8}$ law, with both curves remaining sub-millimeter at 10\,m.

\subsection{Insight~3: ToF Provides Limited Resolution; Coupling Gradient Dominates}
\label{subsec:paradigm}

RF-ISAC extracts range from the round-trip propagation delay, yielding a
resolution limit of $c/(2B)$.
At MI bandwidth $B\!=\!1$\,kHz, this gives 150\,km---approximately six
orders of magnitude coarser than the sub-millimeter MI coupling-based
resolution, rendering ToF-based ranging \emph{unsuitable} at MI
bandwidths.

\begin{figure}[!t]
  \centering
  \includegraphics[width=\columnwidth]{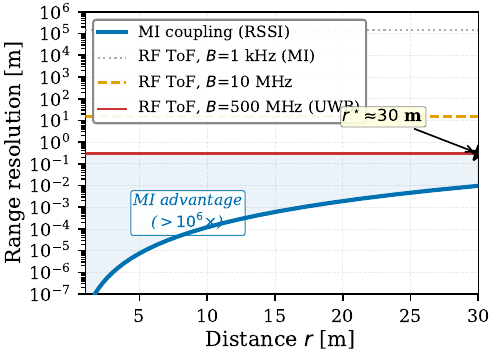}
  \caption{Resolution comparison: MI coupling-based (RSSI) vs.\ RF ToF
    at various bandwidths.
    Within the MI communication range ($r < r^{\star}$), MI achieves
    over three orders of magnitude finer resolution than 500\,MHz UWB.
    The paradigm shift: from bandwidth-limited ToF to gradient-limited
    RSSI.}
  \label{fig:tof_rssi}
\end{figure}

MI-ISAC instead exploits the pronounced
$r^{-3}$ amplitude dependence, which creates a measurable signal change
for sub-millimeter displacements.
This is \emph{gradient-based ranging} rather than \emph{delay-based
ranging}.
The crossover distance $r^{\star}$ below which MI outperforms even
wideband RF (500\,MHz UWB, resolution 0.3\,m) is on the order of
10\,m for representative parameters, comfortably within the typical MI
communication range.
Fig.~\ref{fig:tof_rssi} illustrates this paradigm shift: within the
MI operating region, MI achieves over three orders of magnitude finer resolution than UWB.

\subsection{Insight~4: 4--10+\,dB ISAC Gain over TDMA}
\label{subsec:isac_gain}

To quantify the benefit of \emph{integrating} sensing with communication
rather than separating them in time, MI-ISAC is compared against a TDMA
baseline where a fraction $\alpha$ of the frame is allocated to
dedicated sensing pilots.
MI-ISAC uses the \emph{entire} $N$-symbol frame for both functions,
while TDMA uses only $\alpha N$ symbols for sensing.
The CRB ratio yields a sensing gain inversely proportional to the
pilot fraction $\alpha$ (in dB), which decomposes into
two interpretable terms.

The first is a \emph{time-multiplexing gain}: at $\alpha\!=\!0.5$
(half the frame for sensing), MI-ISAC observes twice as many sensing
symbols, yielding a 3\,dB advantage.
The second is a \emph{structural-information gain} of 1--9\,dB:
the deterministic MI channel enables non-data-aided (NDA) estimation
from \emph{each} received symbol---including data symbols---without additional overhead.
In TDMA, only dedicated pilots contribute to sensing.
At sufficiently high MI-range SNR (tens of dB, typical in the
short-range reactive near field), NDA channel estimation approaches
data-aided performance because the constellation-removal penalty
becomes negligible: each symbol can serve as a sensing pilot.
The structural gain increases with SNR and the number of coil axes,
reaching 6--9\,dB for tri-axial configurations.
For practical sensing allocations ($\alpha = 0.1$--$0.5$), the total
gain is 3--10+\,dB.

\subsection{Insight~5: MI-MIMO---$\mathrm{rank}\!=\!3$ and
  $\kappa\!\equiv\!2$}
\label{subsec:kappa}

The tri-axial MI-MIMO channel inherits the eigenstructure of the
dipole coupling tensor $\vect{G}$.
A direct eigenanalysis yields eigenvalues $\{+2,-1,-1\}$ for all
orientations, giving rank\,=\,3 and a universal condition number
$\kappa\!\equiv\!2$.

The eigenvector for $\lambda\!=\!+2$ is $\hat{\vect{r}}$ itself (the
\emph{radial mode}), representing the strongest coupling along the
Tx--Rx axis with $4\times$ effective power gain.
The two degenerate eigenvectors ($\lambda\!=\!-1$) span the
perpendicular plane (\emph{tangential modes}), coupling to angular
displacements.
This eigenstructure induces a natural ISAC mode decomposition:
allocate power on the radial mode for maximum communication rate,
and use the tangential modes for independent angular sensing.
Because eigenvalues are known \emph{a priori} (they depend only on
direction, not on fading), channel state information feedback is
not required---a distinct advantage over RF-MIMO ISAC, where the condition
number depends on scattering and can be arbitrarily large in
line-of-sight channels.
The eigenstructure itself is an established property of the magnetic
dipole tensor; the contribution here is identifying its role as a
\emph{geometry-independent ISAC mode decomposer} that decouples
range estimation from angular sensing.

\section{Use Cases and Deployment Sketches}
\label{sec:usecases}

The three RF-denied scenarios from Fig.~\ref{fig:concept} are revisited,
mapping MI-ISAC capabilities to concrete operational needs.

\subsection{Underground: IoUT, Mining, and Pipeline Monitoring}
\label{subsec:underground}

Underground wireless sensor networks deployed for pipeline integrity
monitoring, precision agriculture, and mine safety currently use MI
nodes operating at 1--100\,kHz with typical ranges of
5--30\,m~\cite{sun2010magnetic,kisseleff2014throughput}.
MI-ISAC can augment these nodes with simultaneous ranging (inter-node
distance estimation for cooperative localization), soil conductivity
monitoring ($\sigma_m$ from the complex channel coefficient), and link
quality tracking---without additional hardware or spectral resources.
The $r^{8}$ CRB scaling predicts sub-millimeter inter-node ranging at
10\,m in principle, which could complement dedicated ultrasonic displacement sensors
currently used for structural health monitoring.
Temporal monitoring of $\sigma_m$ changes could, if validated, provide early warning of
water pipe leaks or ground saturation preceding landslides.
Relay waveguides~\cite{sun2013deployment} could simultaneously serve as
communication relays and distributed sensing anchors, transforming MI
infrastructure into an underground sensing network.

\subsection{Underwater: AUV Swarm Coordination}
\label{subsec:underwater}

AUV swarms performing coordinated ocean-floor mapping or search-and-rescue
operations require simultaneous communication and relative positioning.
Acoustic ISAC is fundamentally limited by the low
propagation speed of sound (1.5\,km/s vs.\ $3\!\times\!10^{8}$\,m/s),
creating a latency--resolution tradeoff:
centimeter-level acoustic ranging requires $>$10\,ms round-trip times at
10\,m, introducing excessive delays in real-time formation control
loops~\cite{akyildiz2015realizing,li2019survey}.
MI-ISAC provides a low-latency alternative with
coupling-based sensing, where propagation delay is negligible at MI
ranges.
The $\kappa\!\equiv\!2$ property guarantees well-conditioned positioning
regardless of AUV orientation---important for maneuvering
platforms where attitude changes continuously.
Furthermore, the complex MI channel coefficient carries medium
information: the imaginary part of the received coupling depends on
seawater conductivity, potentially enabling real-time salinity and temperature
mapping as an additional capability of inter-vehicle
communication~\cite{gulbahar2012underwater}.

\subsection{In-Body: Implantable Bioelectronic Devices}
\label{subsec:inbody}

Next-generation bioelectronic implants (pacemakers, neural stimulators,
closed-loop drug delivery) require both telemetry to external controllers
and local tissue sensing for diagnostic feedback.
MI-ISAC can enable an inductively coupled link to simultaneously carry
diagnostic data and estimate tissue impedance changes indicative of
edema, infection, or electrode migration~\cite{ma2025survey}.
This extends the biomedical ``theranostics'' concept (therapy $+$
diagnostics) to the communication layer, where the telemetry link itself
can serve as a diagnostic sensor.
Short communication distances ($<$30\,cm) and high coupling at low
frequency make the $r^{8}$ CRB particularly suitable, potentially enabling
micrometer-level resolution for implant micromotion detection.

\section{Open Challenges and Research Roadmap}
\label{sec:challenges}
\begin{figure*}[!t]
  \centering
  \includegraphics[width=\textwidth]{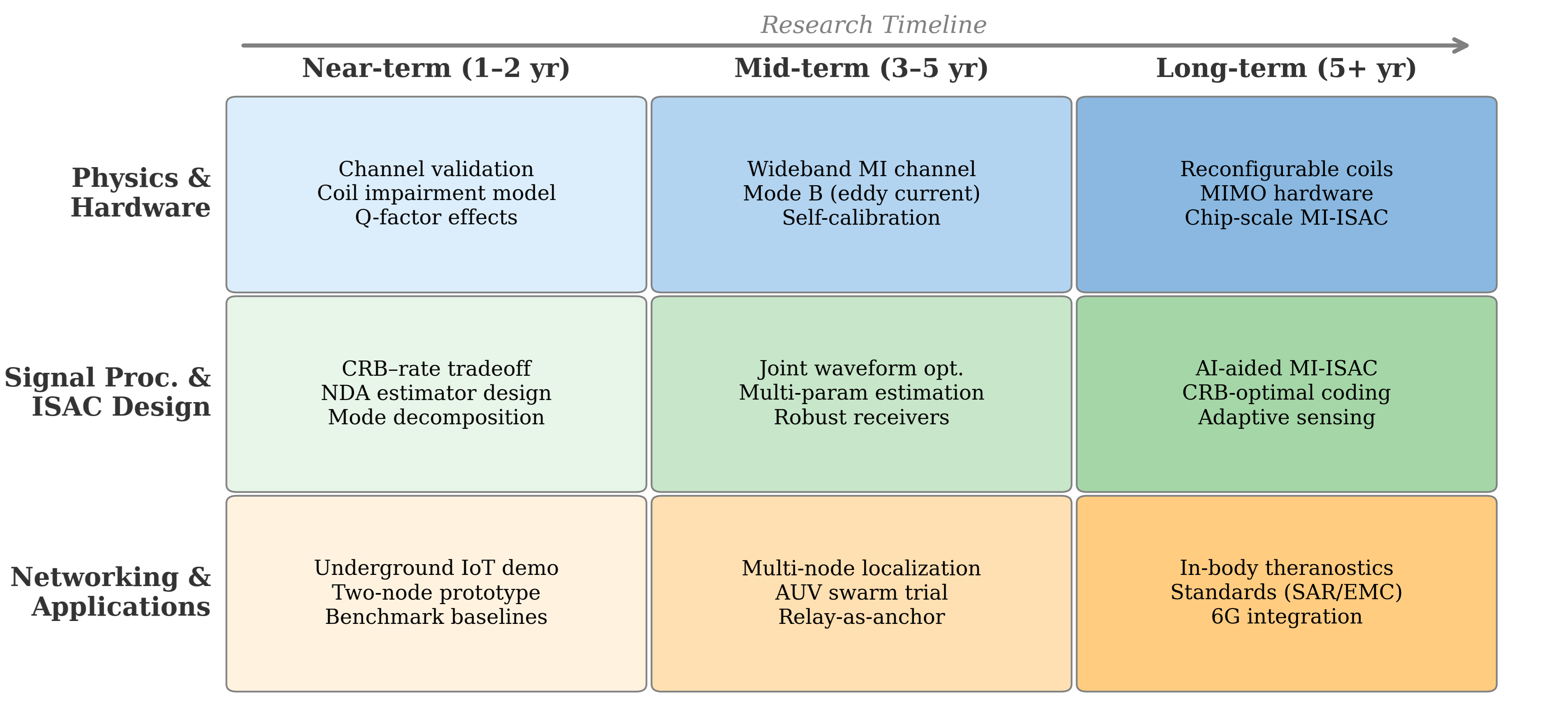}
  \caption{MI-ISAC research roadmap: a $3\!\times\!3$ taxonomy spanning
    Physics \& Hardware, Signal Processing \& ISAC Design, and
    Networking \& Applications across near-term (1--2\,yr), mid-term
    (3--5\,yr), and long-term (5+\,yr) horizons.
    Mode~B (self-impedance sensing) is a future extension.}
  \label{fig:roadmap}
\end{figure*}

The open research landscape is organized into four challenge areas
spanning physics, signal processing, and networking, and summarized in
the roadmap of Fig.~\ref{fig:roadmap}.

\textbf{1) Model Validity and Hardware Non-Idealities.}
The results in this article rely on the magnetic dipole approximation
($a \ll r$), ideal coil alignment knowledge, and thermal-noise-limited
reception.
Practical MI transceivers face coil impedance drift with temperature,
finite Q-factor limiting bandwidth, power amplifier nonlinearity, and
ADC quantization noise.
Fig.~\ref{fig:crb_fim}(b) provides a first-order assessment: a
combined 9\,dB front-end penalty shifts the CRB upward but preserves
the $r^{8}$ scaling law.
A systematic investigation of these impairments on the
$\kappa\!=\!2$ invariant, and self-calibration protocols exploiting
tri-axial redundancy (9 measurements for 3~unknowns), are potential
near-term directions.

\textbf{2) Multi-Node Interference and Scheduling.}
In networks of $K$ MI-ISAC nodes, co-frequency mutual coupling creates
inter-link interference degrading both functions. 
MI interference decays with the same $r^{-3}$ order as the desired coupling; 
hence the signal-to-interference ratio does not improve with distance, 
and interference must be explicitly managed through scheduling or spatial separation.

\textbf{3) ISAC Waveform and CRB--Rate Tradeoff.}
The RF-ISAC literature on CRB--rate region characterization and
dual-function waveform design~\cite{liu2022fundamental} provides an
established analytical toolkit, but the underlying physics need to be
adapted for MI.
The MI-ISAC Pareto boundary between communication rate and sensing
Fisher information admits a closed-form expression because the channel
is deterministic, enabling globally optimal CRB--rate tradeoff designs
that are intractable in stochastic RF channels.
Mode decomposition (radial for communication, tangential for sensing)
provides a natural starting point for waveform co-design.

\textbf{4) Cooperative Localization, Benchmarking, and Field Trials.}
Fusing range/angle estimates from $K$ links enables anchor-free 3D
localization, where $\kappa\!\equiv\!2$ simplifies GDOP analysis.
Fair benchmarking requires standardized baselines (TDMA-MI, pilot-only
MI, RF near-field ISAC) and experimental validation in soil, seawater,
and tissue phantoms.

\section{Conclusion}
\label{sec:conclusion}

This article has introduced MI-ISAC---a reactive near-field ISAC paradigm
that redefines the sensing mechanism from bandwidth-limited
time-of-flight to coupling-gradient-based ranging.
Three key results characterize the contribution:
(i)~tri-axial coils are necessary and sufficient for identifiable
sensing;
(ii)~the CRB scales as $r^{8}$, with a theoretical limit of sub-millimeter accuracy at
typical MI distances; and
(iii)~the MI-MIMO coupling tensor has a universal $\kappa\!=\!2$ and
rank\,=\,3, providing well-conditioned sensing for all geometries.
MI-ISAC can achieve 4--10+\,dB sensing gain over time-division baselines,
making sensing essentially free in high-SNR MI links.
MI-ISAC can serve as a foundational building block for 6G connectivity
in challenging environments---underground, underwater, and
inside the human body---opening a broad design space for future
investigation.

\bibliographystyle{IEEEtran}

\bibliography{references}

\vfill

\begin{IEEEbiography}
[{\includegraphics[width=1in,height=1.25in,clip]{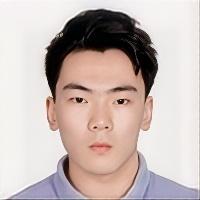}}]{Haofan Dong (hd489@cam.ac.uk)}
 is a Ph.D. student in the Internet of Everything (IoE) Group, Department of Engineering, University of Cambridge, UK. He received his MRes from CEPS CDT based in UCL in 2023. His research interests include integrated sensing and communication (ISAC), space communications, and THz communications.
\end{IEEEbiography}

\begin{IEEEbiography}[{\includegraphics[width=1in,height=1.25in,clip]{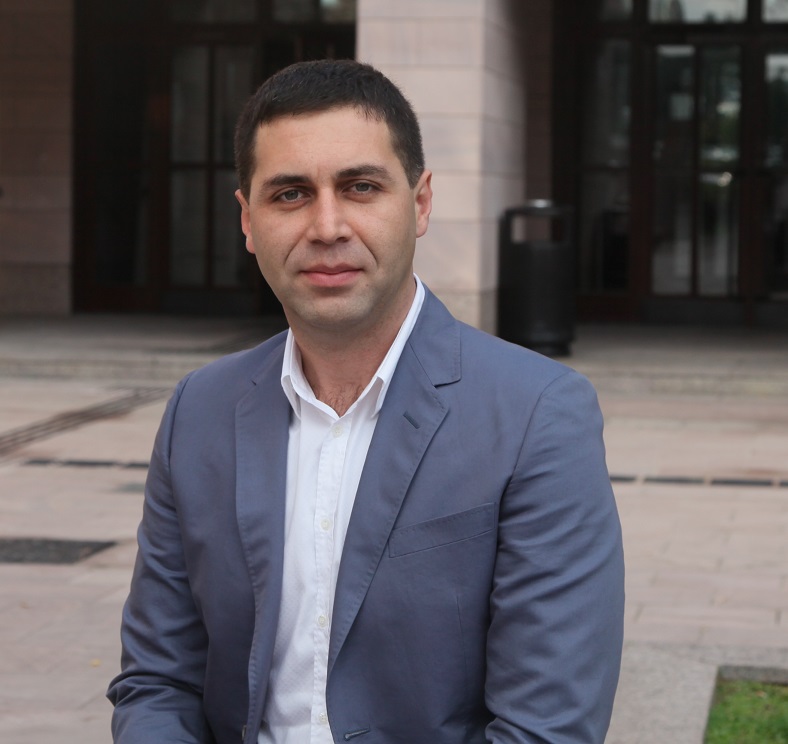}}]{Ozgur B. Akan (oba21@cam.ac.uk)}
received his Ph.D. degree from the School of Electrical and Computer Engineering, Georgia Institute of Technology, Atlanta, in 2004. He is currently the Head of the Internet of Everything (IoE) Group, Department of Engineering, University of Cambridge, and the Director of the Centre for NeXt-Generation Communications (CXC), Koç University. His research interests include wireless, nano-, and molecular communications, and the Internet of Everything.
\end{IEEEbiography}

\end{document}